**Seamlessly Integrating Loops That Matter into Model Development and Analysis**
By William Schoenberg and Robert Eberlein

**Abstract**
Understanding why models behave the way they do is critical to learning from them, and to conveying the insights they offer to a broad audience. The Loops that Matter methodology automatically shows which loops are dominating behavior at each point in time and generates simplified causal loop diagrams from a user adjustable set of important loops. This paper describes the challenges of implementing these tools into a fully functioning model development environment along with the solutions developed. The promise of the tools has, if anything, been amplified by the results of this implementation, and we give several examples of using the tools. For pedagogical models Loops that Matter can ease communication while speeding and deepening learning. For complex models the tools allow the extraction of realistic explanations of behavior in the form of animated simplified causal loop diagrams. For models with discrete and discontinuous elements, the bigger feedback picture is still easily discoverable. While there will doubtless be refinements and enhancement to the delivered tools, they represent a large step forward in our ability to understand models from conceptualization through delivery.

**Introduction**
Automatically discovering the origins of behavior within real-world models is a 'holy grail' pursuits for the field of system dynamics. Over the past 40 years, dozens of researchers have dedicated many hours and publications to this task (see for example: Graham, 1977; Forrester 1982; Eberlein, 1984; Davidsen, 1991; Mojtahedzadeh, 1996; Ford, 1999; Saleh, 2002; Mojtahedzadeh et al., 2004; Güneralp, 2006; Gonçalves, 2009; Saleh et al., 2010, Kampmann, 2012; Hayward and Boswell, 2014; Moxnes and Davidsen, 2016; Oliva, 2016; Sato, 2016; and Hayward and Roach, 2017). All of these approaches offer a wealth of benefits, some unique to an approach, but many shared across approaches. To date, none of these approaches are in common use by a significant number of modelers, or students of modeling. Some of this is due to deficiencies, as each approach does have systematic limitations and blind spots (Kampmann & Oliva, 2009), but more importantly all of them require the practitioner to do significant work. People with years of training, education, and experience, tend to rely on the understanding they develop as they build and work with models. Those with less experience are overwhelmed by the challenges of building good models and don't have the capacity to learn a totally new toolset at the same time.

All of this is prelude to saying that if tools for discovering the origins of behavior are simply part of the model development experience, they will, in turn, be part of how people understand the models they build an work with. For experienced practitioners they will both reinforce and challenge beliefs about what is truly driving behavior. For those with less experience they will offer pathways to understanding that are both discoverable and easy to communicate. That implementation, and the way it can be used for discovery and communication, is what we are presenting in this paper.

Starting with Version 2.0 of the Stella software products (including Professional, Architect, and eventually the online tools), the Loops that Matter method (Schoenberg et. al, 2019) are simply part of the modeling experience. It is turned on by a checkbox, and automatically reports information about loops and links that matter and allows visual exploration of the model structure crucial to driving behavior.

This paper deals with the implementation challenges and outcomes of the Loops that Matter method, not the method itself. Nonetheless, it is useful to provide some background on the different approaches there are to understanding behavior analytically, and where the method fits in. It shares many of the characteristics, and benefits of existing work, but has an advantage in computational simplicity. For completeness, we will then quickly define the link and loop scores that form the backbone of the method. This is followed by discussion of a number of implementation challenges and opportunities, and finally a number of example applications to well known and pedagogical models.

**Existing Approaches to Loop Dominance Analysis**
Analytic methods for determining loop dominance have historically revolved around two approaches. The first is based on eigenvalue elasticity analysis (EEA), the second uses the pathway participation metric (PPM) and causal pathways.

EEA is used to determine what combination of behavior modes a given model structure produces (Saleh, 2002; Kampmann et al., 2006; Saleh et al., 2010; Oliva, 2016). EEA uses a linearization of the model and its associated eigenvalues and eigenvectors as the unit of analysis. It is the most encompassing method for structural analysis of models, but is limited in the set of the models it can analyze without modification due to its toolset (Oliva, 2016). Much work has been done to make tools supporting EEA easier to use, and usable on models with a wider variety of formulations, but it is fundamentally designed to work on continuously differentiable systems. Current software tools for performing EEA change model equations to meet that requirement which has measurable, impacts on simulation results (Oliva, 2016).

Current applications of EEA are based on an independent loopset as described by Kampmann (2012), or the unique shortest independent loopset as described by Oliva (2004). The independent loopset (and its variants) provides a reasonable number of loops to analyze and also maintains loop independence which many tools using the EEA approach rely upon. However, the use of an independent loopset (and its variants) can also limit the ability to understand classes of models where behavior is fundamentally driven by loops outside of these sets as described by Eberlein and Schoenberg (2020).

PPM, unlike EEA, does not use eigenvalues to describe model structure, instead it focuses on the links between variables, specifically tracing the causal pathways between stocks, and identifying the causal pathway most responsible for moving the stock in the direction of its net change (Mojtahedzadeh et al, 2004). PPM based methods (including Hayward and Boswell's Loop Impact Method (2014)) study observed behavior modes relative to individual stocks in the model, rather than all stocks together. The PPM toolset is directly applicable to more models because it does not require continuously differentiable systems, but it has been criticized for its failure to clearly explain oscillatory behavior (Kampmann & Oliva, 2009)

**The Loops that Matter Method**
The Loops that Matter method uses computations based on actual variable values during simulation. As a consequence, it can be applied to models of any size or complexity without regard to continuity. It produces analyses which match those done by PPM and EEA when applied to the same models (Schoenberg et. al, 2019). The method, like PPM, builds upon observations of the way that experienced modelers perform analysis to determine the sources of observed behavior. The method does all calculations directly on the model equations, making it easier to understand the measurements of loop and link contributions to behavior. The method uses only values computed as part of the normal simulation of the model, making it applicable to models with discrete characteristics.

The Loops that Matter method produces three key metrics, the 'link score' for any link in the model, and 'loop score' as well as 'relative loop score' for any loop in the model. Together these metrics provide the necessary information to compute, visualize and animate the origins of behavior in system dynamics models. A full discussion of the Loops that Matter method is contained in Schoenberg et. al, 2019, but a brief summary of the method is provided below for convenience.

**Summary of the Link Score**
The link score (Equation 1) is best thought of as a contribution at one instant in time from an independent variable $x$ to a dependent variable $z$ or as a measure of the contribution and the polarity of the link from $x$ to $z$. The link score for the link x → z is:

*Equation 1: The discrete form for the link score equation which matches the implementation of the calculation and is computed each dt.*

$$LS(x \to z) = \begin{cases} \left(\left|\frac{\Delta_x z}{\Delta z}\right| \cdot sign\left(\frac{\Delta_x z}{\Delta x}\right)\right), \\ 0, \quad \Delta z = 0 \text{ or } \Delta x = 0 \end{cases}$$

In Equation 1 Δz is the change in z from the previous time to the current time. Δx is the change in $x$ over that interval. $\Delta_x z$ is the change in $z$ with respect to $x$ over that interval. From a computational perspective $\Delta_x z$ which is called the partial change in $z$ with respect to $x$, is the amount $z$ would have changed, conditionally, if $x$ had changed the amount it did, but $y$ had not changed. The first major term in Equation 1 represents the magnitude of the link score, the second is the link score polarity as defined by Richardson (1995).

Because stocks represent an integration process, they change over time as a result of the values of flows, not because of changes in those flow values. For this reason, the computation of link scores for links representing the effects of flows on stocks is different. A deeper discussion of the construction of this equation can be found in Schoenberg et., al. 2019. Assume the stock equation $s = \int(i - o)$ where $s$ is the stock, $i$ is the inflow, and $o$ is the outflow. The link score in these cases is computed as seen in Equation 2.

*Equation 2: Link score for all links from flows to stocks (both inflows and outflows are covered)*

$$Inflow: LS(i \rightarrow s) = \left(\left|\frac{i}{i-o}\right| * 1\right) \quad Outflow: LS(o \rightarrow s) = \left(\left|\frac{o}{i-o}\right| * -1\right)$$

One of the important attributes of link scores is that they can be multiplied together to give the effect along a path between an input and an output. This product is called the path score, and used to deal with hidden paths, such as those that arise from macros as discussed below.

**Summary of the Loop Score and the Relative Loop Score**
The loop score is computed by multiplying all link scores together for each link in a loop at a point in time. This is the same as the path score from a variable back to itself, because a loop is simply a closed path. The magnitude of the loop score represents the contribution of a loop at a time to changes in all model variables[1]. Loop scores can be very large in magnitude, so we typically use the relative loop score, a normalized value, computed so that the absolute value of all relative loop scores add to 100%. The relative loop score measures the percentage contribution a loop to the changes of all variables in the model[1] at each point in time.

**Challenges to making Loops that Matter production ready**
As should be clear from the preface, our goal in making Loops that Matter a part of Stella was to have something seamlessly integrated and, from the user's perspective, essentially free to use. It is a truism of software development that the easier you make a feature for users, the more the effort required to develop the feature. This is an expected part of all commercial software development. A great deal of effort went into making both the computational engine, and the user interface as efficient and effective as possible. There were, in addition, a number of obstacles to having the software simply work that are more conceptual. Some require extensions to, or at least new interpretation of, the Loops that Matter method. Some are decisions on how to best present information. All effect the results, as seen by the user, and so are worthy of discussion.

**Macros**
The first challenge to overcome was how to handle macros such as DELAY or SMOOTH which incorporate a complex hidden internal structure. Because of the hidden structure, links that appear to the practitioner as directly connected on the diagram, are often times quite indirect which is demonstrated in the example below. This problem is especially complex because there may be multiple causal pathways with differing strengths and potentially even polarities between the input and output of what may appear to the practitioner to be a simple link. In addition, there may even be feedback loops within the macro equations themselves.

---

[1] Assuming all variables share the same feedback loops. For models with feedback loops that exist in isolation (either completely disconnected set of variables, or as inputs to, or outputs from, other parts of the model) the variables are broken into sets that share feedback loops, and the scores computed on each set.

Take for example the DELAY3 macro pictured below in Figure 1. From the perspective of the practitioner there is a direct link between 'input' and 'output using macro', (with a second connection from 'delay time'. When the full set of relationships underlying the macro is shown, however, we see a much less direct path from 'input' to 'output using full structure' even though they are identical computationally. Once the full structure of the macro is exposed it's obvious that there are multiple distinct pathways from 'delay time' to the output, in fact there are six distinct causal pathways if we include the influences to the flows both directly, and through the upstream stocks. For the 'input', there is only a single pathway, but it goes through every stock in the chain. Therefore, what appears to be two simple direct links on the diagram is actually seven distinct causal pathways all but one (the connection of 'delay time' to 'Flow 4') involving intervening stocks. In addition, the structure itself contains three feedback loops; fortunately for this, and most the built in macros, the internal feedback loops do not cause any behavior by themselves.

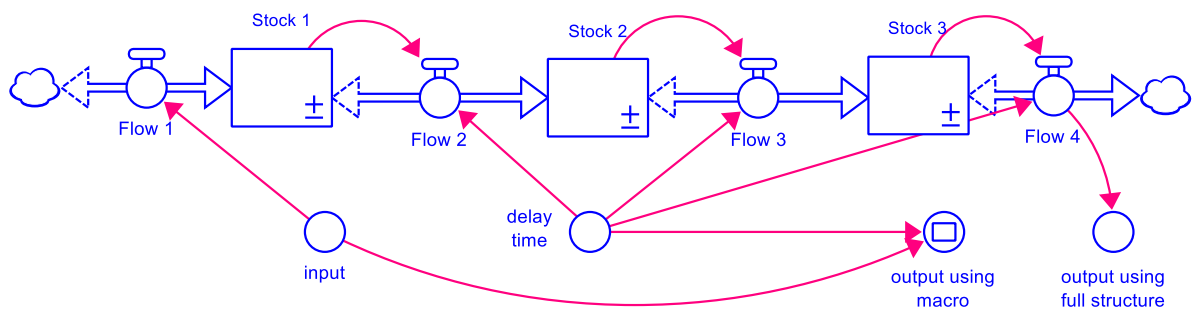

*Figure 1: The structure of the DELAY3 macro, demonstrating the complex set of pathways between the arguments to the macro (input, and delay time) to the output.*

The solution devised to this challenge is a simple heuristic applied at each calculation interval. The link score of a pathway which passes through a macro is the path score of the expanded pathway. If there are multiple pathways, we choose the path score with the largest magnitude (positive or negative). The reasoning behind this definition is that it will maintain the integrity of the loop scores computed through the macros. If there is a single path through the macro the loop score will be exactly what it would have been if the macro had been expanded. If there are multiple paths (so multiple loops) the loop score will be the biggest of all the loops involving the macro. It is also possible, using this definition, to compute the link score as the simulation occurs with no post processing, supporting the definition of a PATHSCORE builtin function that can be used in the model. The link sore for anything going into a macro is, thus, a composite. As a consequence, the structure represented by a link through a macro is not necessarily fixed throughout the course of the simulation run.

We considered, but ultimately rejected, an alternative way of dealing with macros that involved the determination of the dominant pathways through a macro by post-processing all of the path scores and picking the best one. This has the advantage that the structure within the macro is invariant, but would change loop scores relative to the case where the macro were expanded. A similar approach in which macros were given predefined pathways for link score computation was also rejected for similar reasons. Another approach considered was the expansion of all macros with the internal information directly exposed to the end user. This solution was rejected because it would be confusing to the practitioner and the macro variables would have to be given names based on usage which would be hard to read and be detrimental to the quality of simplified CLDs, which will be discussed later. In addition, expanding the macros would ignore one of the main reasons for using them, which is to prevent the clutter of unnecessary detail in visual representations.

The link score we use for macros (called the composite link score) does preserve overall loop scoring, but macros will still look different from their expanded versions. For example, if the DELAY3 macro is given a step input (and nothing else), the reported link score will always be 0. This is because the input is changing only at a single point in time, and the output is not changing at that time. Once the output starts to change, the input is no longer changing. Since the reported link score is the product of the internal link scores, one of which is always 0, the composite score is 0. This makes macros somewhat inscrutable when they are not actually part of any feedback structure.

One other point worth noting on macros, is that the loops involving internal macro variables need to be trimmed of those internal variables before being reported. Similarly, loops internal to the macro (such as the first order drains in the DELAY3 example) are dropped altogether and not reported.

**Discrete Variables and Stateful Functions**
Stella includes a number of discrete elements (Conveyors, Queues and Ovens) as well as builtin functions like PREVIOUS that retain state values. We won't go into details here, as many of these have been dealt with on a case by case basis, but will just point out the principals by which we have dealt with them. Unlike macros, for which there is a rigorous solution to following paths, there are internal structures in the elements and functions that can't practically be exposed to complete link score computation. A conveyor, for example, can have thousands of individual elements waiting to be used at a later time. To compensate for this, we have approximated an instantaneous response to changing inputs based on the perfect mixing analog in traditional System Dynamics models.

Consider, for example, a conveyor. If the input changes nothing happens till much later in the simulation. This is distinct from a normal stock with a proportional outflow where a change in the stock value causes the outflow immediately changes. For the conveyor the change is not immediate, but eventual, although it still has the same basic character. So we treat conveyors as if the instantaneous response is the eventual response. This may cause some distortion in the time profile of the loop dominance, but it gets polarity and magnitude correct. It also, in practice, seems to work quite well as can be seen from one of the examples we discuss.

**Too Many Loops**

Feedback is pervasive, and some models have more feedback loops than can be dealt with in reasonable time by the software, let alone a practitioner. The solution to this problem is to focus only on the loops that matter based on their relative loop scores. That solves the problem for the practitioner, but not for the computer in the cases where models have more than a few thousand loops. Since the number of potential loops is typically more than linearly proportional to the number of stocks in a model, it can get very large very fast (Kampmann, 2012). How big it actually gets depends on the model, but a number of published models have too many loops to practically enumerate. As stated above, there are static analysis techniques for dealing with this problem, namely the independent loopset and the shortest independent loopset as used by EEA. Static analysis, unfortunately, does not allow the loops under consideration to change over the course of a simulation, which is exactly what is required as dominance shifts, and can therefore miss important feedback as demonstrated by Güneralp (2006) and Huang et. al, (2012). The strongest path algorithm described in Eberlein and Schoenberg (2020) solves this problem with a heuristic that finds the loops that matter at each point in time, with the resulting set of loops used to do further analysis.

The strongest path algorithm can change the loops that are identified based upon the parameterization of a simulation run. Just as the most important loops change with different parameters, so will the loops actually identified for large models. This is problematic if the intent is to study the impact of a loop across a wide set of input parameters, since the loop under study may not always even be identified. To address this problem, a LOOPSCORE builtin was devised which allows the practitioner to specify any arbitrary feedback loop, and compute its score over time. Inclusion of this guarantees that the loop will be reported on regardless of its importance based on the selected input space of the model. This also allows comparisons across runs to see how active a loop is under different scenarios. Loop scores are reported using the LOOOPSCORE builtin function but (because they reported as relative values) they are only available at the end of the simulation. It is, however, possible to compute the raw loop score using the PATHSCORE builtin previously described.

**Visualization of the Loops that Matter**
The final chapter of *Business Dynamics* (Sterman, 2000) makes clear its call for software that can perform "Automated identification of dominant loops and feedback structure", while "Linking behavior to generative structure" and performing "Visualization of model behavior". The LoopX tool developed by Schoenberg (2019) was the first of its kind to make major headway on these lofty aspirations. Using the link score, LoopX was able to create animated stock and flow diagrams where the connectors and flows change size and color relative to the magnitude and polarity of the link scores. Further still, LoopX demonstrated how to use the loop score in combination with the link score to quickly and dynamically simplify model presentation by automatically generating high quality animated causal loop diagrams, selecting only those variables and links which were absolutely necessary to explain the feedback behavior of the model at the requested level of complexity. That technology has been refined in the current implementation, and its scope expanded to encompass practitioner developed models with high levels of complexity.

The LoopX tool introduces two important terms, the link inclusion threshold and the loop inclusion threshold, both as a part of the CLD simplification process. The simplification process suggested by Schoenberg (2019) filters the list of variables in the simplified visualization by specifying which variables should be kept based upon the variation in the magnitude of the relative link score across the entire simulation period. Only variables with an inbound link (causal connection) with a relative link score that varies by at least the link inclusion threshold will be kept. The loop inclusion threshold keeps the stocks (and optionally flows) in every loop that explains, on average, at least the specified percent of model behavior. Since all loop scores are presented as relative values with magnitude adding to 100, this is straightforward to determine.

From an implementation perspective, the main problem with the methods laid out in the LoopX tool was that the variables to include would first be selected, then all pathways connecting those variables determined, and then the loops of the simplified diagram detected. This approach has some computational shortcomings, as loop detection is not always easy to do, and some conceptual shortcomings as it is difficult to relate the simplified loops to the original loops and it is quite likely that unimportant loops will be included in the reported set of loops.

The implementation we developed uses the same metrics for the selection of variables to include, but then uses the original loops to create the simplified diagram. This way there is a two way mapping from the loops in the full diagram to the loops in the simplified diagram (though more than one loop in the original model may result in the same loop in the simplified diagram). This makes it possible to compute metrics such as the fraction of total model behavior explained by the simplified CLD as well as to attach scores to the simplified CLD relative to the original model. This also means that the connections to be included in the simplified diagram are the important ones. We do, however, perform a small amount of loop closure of the simplified diagram to improve the layout, and this can add some unimportant (though aesthetically pleasing) loops to the simplified CLD.

Because the source of all the feedback loops in the simplified CLD is known, each can be given a composite relative loop score. This is the sum of the relative loop scores from all full loops which reduce to that simplified loop. Since relative loop scores are normalized, and each full loop is only marked as being represented by a single simplified feedback loop, the sum of the composite relative loop scores for all loops in a simplified CLD represents the portion of the full behavior of the model explainable by the simplified CLD. This measure proxies the quality of the simplified CLD. Numbers closer to 100 mean that more of the model's behavior is being explained by that simplified CLD, closer to 0 mean that less behavior is represented in the presented simplified CLD.

A surprising outcome of deriving simplified loops from the full loops is that it is possible to produce simplified CLDs where the loops are disconnected. While at first glance this outcome may seem problematic, upon reflection it is representative of the feedback structure of the underlying model. Take for example Figure 2, where a model is shown which has two strong minor feedback loops that are tied together by a weak major feedback loop. For this sample model, it makes sense that when the loop or link inclusion threshold are set high to achieve a simple CLD, the weak major loop (and its links) disappear, leaving two unconnected minor feedback loops as the primary drivers of model behavior.

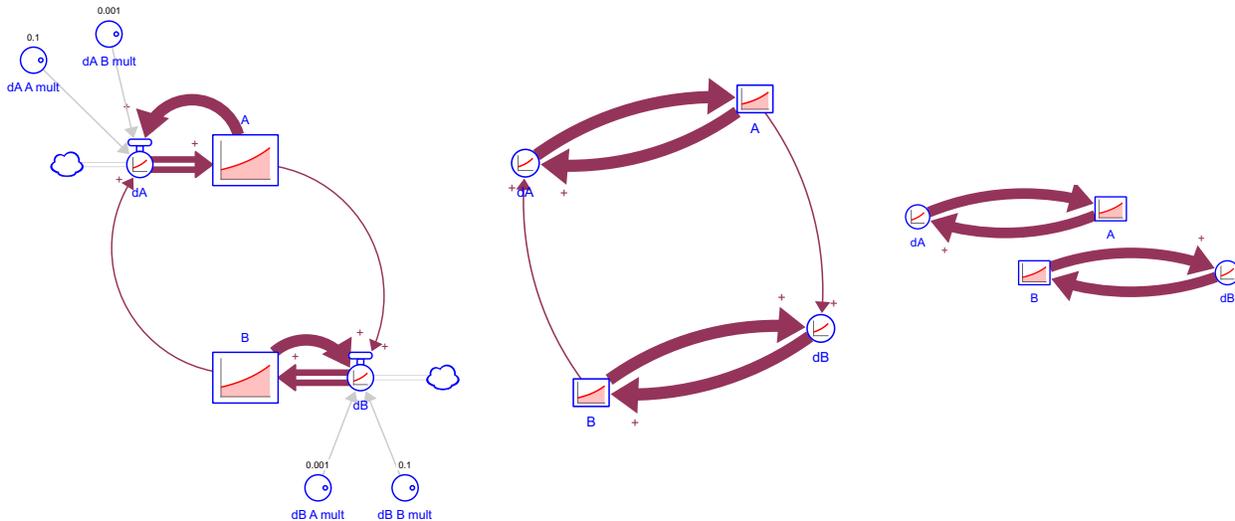

*Figure 2: Demonstration of a weakly coupled system which when simplified turns into two disconnected systems. The far left is the Stock Flow Diagram for this system. The middle is a CLD showing all feedback relationships in the model. The far right shows the most important feedback relationships in the model.*

**Link Thickness and Polarity Markings**

The second challenge in the visualization of simplified CLDs choosing the thickness and polarity for a simplified link. These links may represent a several different causal pathways with different strengths, and potentially polarities. This is similar to the problem of representing composite link scores in macros, but has a different solution. In this case, we use the causal pathway that has the largest path score magnitude averaged across time. The reasoning behind this selection is that the simplified CLD is based on selections of links and loops using their average strength, so it is logical to use a consistent measure for the representation of the link. If a variable is included because it has a strong link into it, then the strength of that link should be presented. Similarly, if a stock or flow is included because of a strong loop, the strength around that loop should be represented. This is a satisfying argument for links which represent causal pathways of the same polarity. For simplified links which have been over-abstracted, where they represent pathways of both reinforcing and balancing polarities, this solution produces simplified CLDs which may be misleading, and we have developed a measure to highlight that situation.

For all simplified links, a confidence value is generated using Equation 3 below, where $r$ is the sum of the single highest magnitude instantaneous reinforcing pathway scores across the entire simulation and $b$ is the sum of the single highest magnitude instantaneous balancing pathway scores across the entire simulation.

*Equation 3: Equation for computing the confidence that a simplified link has a single and consistent polarity*

$$confidence = |(r - |b|)| / (r + |b|)$$

This confidence value makes it very clear when a simplified link is representing two pathways of different polarities (it is 1 if either r or b is 0). A confidence value of 0.99 (or lower) was chosen as the cutoff point which is used to change the color representing the polarity of the simplified link to gray (to represent mixed polarity) which makes it abundantly clear that the simplified CLD is over-simplified. The gray color represents that the two variables are related, but that a key portion of the relationship is omitted preventing a proper polarity of the relationship from being displayed. Figure 3 demonstrates this problem in a simplified CLD representing the dynamics of the base case of Forrester's (1968) market growth model.

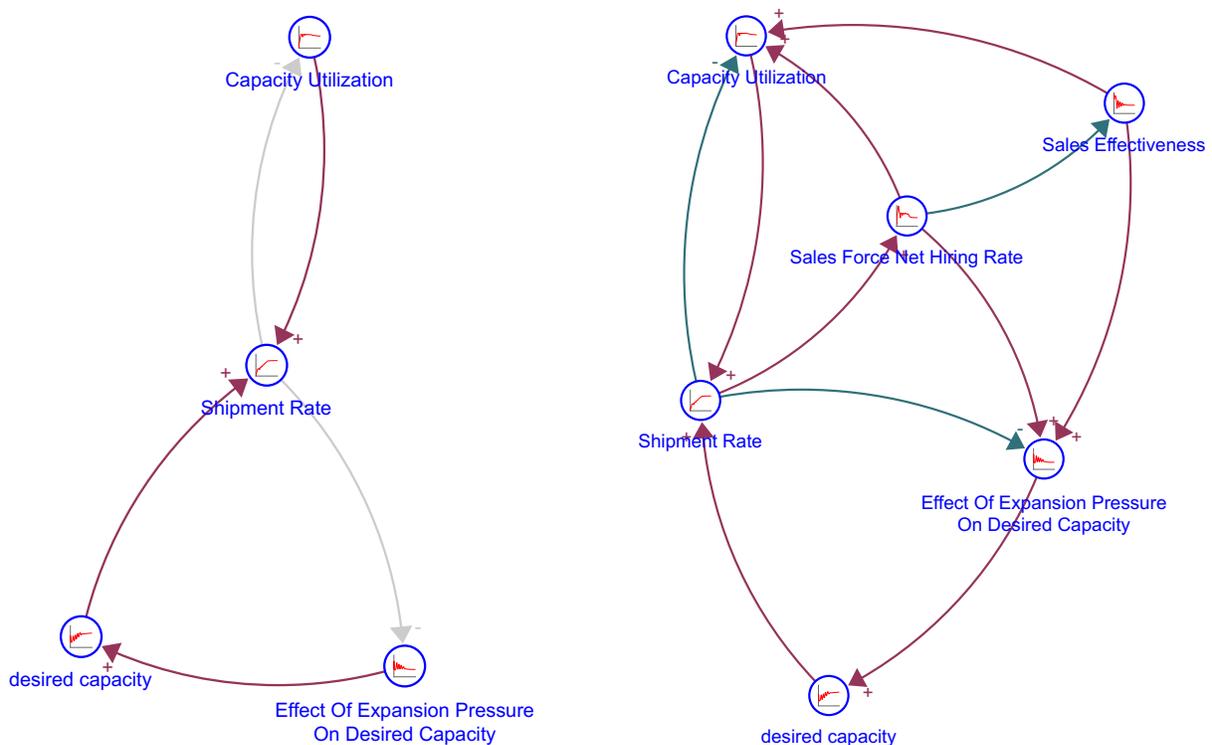

*Figure 3: On the left is an oversimplified version of Forrester's 1968 market growth model with a gray link, which clearly communicates that the link does not have a determinable polarity. The simplified CLD on the right is after adjusting the link inclusion threshold down by one one-hundredth of a percent, it demonstrates how an over oversimplified link is expanded to its key constituent pathways*

**Loops with Unknown Polarity**

Some models contain links, and therefore loops, that change polarity during a simulation. For example, in the yeast alcohol model examined by Schoenberg et. al (2019) there is a link from yeast concentration to growth that is at first positive, then negative. While such links are rare, only occurring in equations that compound multiple effects, they still occur with enough frequency that they have to be dealt with in a production ready system. The problem with these links is that they make interpreting the polarity of a loop across the full simulation time technically impossible because any loop including them most likely has expressed both positive and negative polarities at different points in the simulation.  Attempting to discover the origins of behavior in a model where one of these links is along a key causal pathway, shared by many of the key loops in the model, is very difficult unless there is some way to classify the actual behavior of the loops these links are a part of.  To solve this problem a predominate polarity needs to be established if possible.  Loops are labeled according to the following scheme, Rx, Bx, Rux, Bux, Ux, where x is the index of the loop of that type. R for reinforcing, B for balancing and U for unknown polarity. Ru means unknown polarity, predominantly reinforcing, and Bu means unknown polarity predominantly balancing.  The Ru and Bu designations are assigned when the confidence value for a loop polarity is above .99, as calculated using Equation 3.  This cutoff allows for a well-reasoned factual interpretation of full and simplified CLDs including these links, where the polarity changing nature of these links is not important over the course of the simulation.

**An Additional Option to Simplify CLDs**
The final improvement to the visualization of simplified CLDs is targeted at models with many stocks, and long loops.  The original definition of the link inclusion and loop inclusion threshold state that anytime a stock is kept, so are its flows, regardless of the relative link scores of those flows.  In large models with large numbers of stocks, especially where the feedback loops tend to be long, we get extra flows, which do not necessarily add anything to understanding.  Therefore, a third simplification parameter, a boolean, was introduced allowing the user to control if flows are automatically kept if a stock is kept.

**Demonstration of the Power of the Approach**
Building intuitive understanding of the patterns of structural dominance in practitioner-built models is straightforward using the tools that are now part of Stella.  To demonstrate this, we show three examples. This first is pedagogical, and shows how loop identification and highlighting on the stock and flow diagram can help students understand simple models. The second example is from a relatively complex model that shows the ability of the approach to give insight into larger models. The final example is from a model using a conveyor and a stock with a non-negativity constraint and demonstrates the robustness of the approach to nonlinearity and special conditions.

**Pedagogy**
All of the loop dominance work, including Loops that Matter, was inspired by the desire to understand things that weren't simply obvious. Simply obvious, however, means something very different to an individual who has never built a model than an experienced System Dynamics modeler. After implementing these tools in Stella, and then just using the software in day to day activities with the tools (by default) turned on, it became clear they could help with even very little things. Things that are obvious, but take too many words to explain to those for who would disagree with that claim.

Start with the simplest population model – Population, births and a birth rate as shown in Figure 4. There is one reinforcing loop, it accounts for all the behavior, and the model displays exponential growth. Student can certainly play with the growth rate at this point to get a feel for exponential growth.

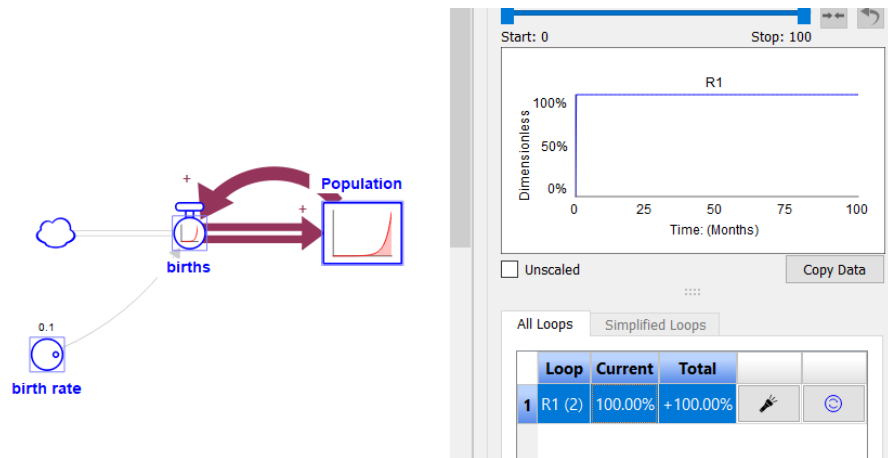

Figure 4: A simple population model with only Births.

Now add deaths simply assuming an average lifetime of 20. The resulting model is shown in Figure 5 and still exhibits exponential growth, but at a much slower rate. The reinforcing loop has a 67% contribution, the balancing loop 35%. If the students shorten the lifetime they will see the percentages move toward 50/-50. If they go less than 10 the graph will show decline and the balancing loop will be dominant.

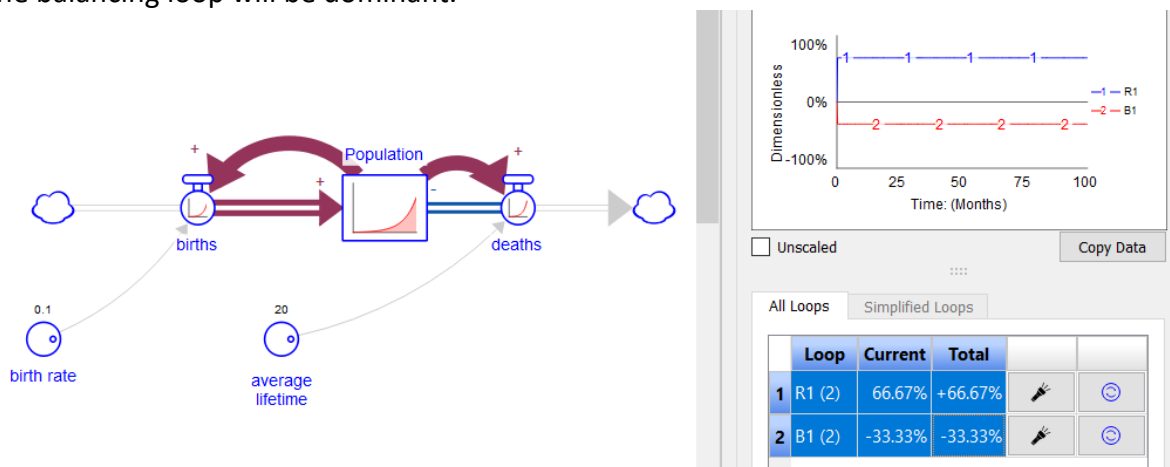

Figure 5: Population model showing both births and deaths

Clicking on the 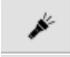 for B1 is also a good way to highlight a loop that does not look like a loop on a stock and flow diagram as shown in Figure 6. This kind of loop, where the outflow from the stock does not have an arrowhead, is harder to students to see.

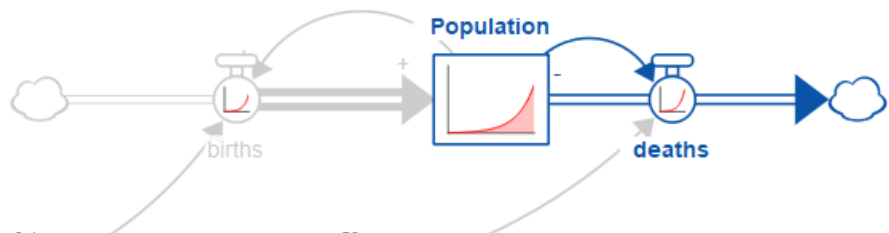
*Figure 6 Highlighting the negative loop involving 'deaths'.*

If the students set 'average lifetime' to exactly 10, then nothing is changing and there are no loops reported. This is a learning moment, both for the students and the teacher. On the one hand the lack of any loops being reported is simply an artifact of the way the Loops that Matter method works. On the other hand, the lack of change is the result of two opposing loops just happening to match. That is either cosmic coincidence, or there is a reason for it, and this allows the introduction of the concept of carrying capacity as shown in Figure 7

This dramatically shows the difference between a fragile equilibrium, and one that is the result of shifting loop dominance. At the end there is one reinforcing loop with a score of 50%, and two balancing loops with scores that add to -50%. The capacity constraint loop (B1) is at first inactive, but then comes to be the bigger of the two balancing loops. The question of whether B1 is changing behavior, or is simply making B2 strong enough to balance R1 should make for good class discussion, though it seems unlikely the class will reach a conclusion.

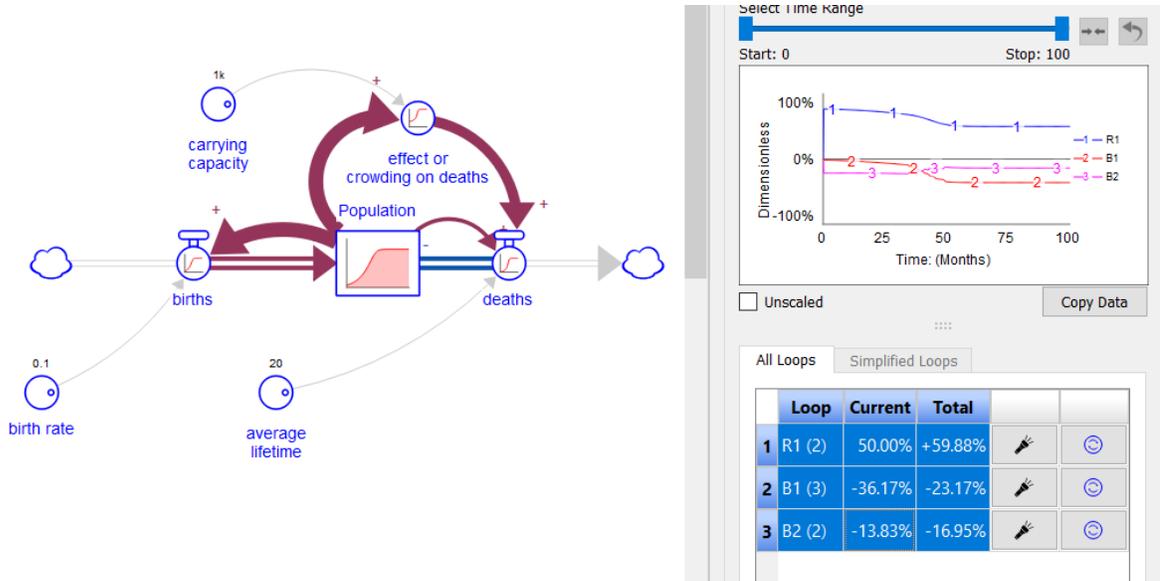
*Figure 7: Population model with the addition of 'carrying capacity'*

This is an extremely simple example, and likely one that many of the readers have used. It demonstrates the extra cues that the Loops that Matter techniques add to the conversation. This has the potential to make the discussion faster, more informative, and better remembered.

**Understanding Economic Cycles**

Mass' 1975 Economic Cycles model provides a good example of a practitioner developed model with a high level of complexity. The model contains 163 variables with 17 stocks, producing 494 feedback loops[2]. Figure 8 shows a machine generated CLD of this model whose 9 simplified feedback loops represent the combined effects of 21 full feedback loops. The 9 simplified feedback loops of Figure 4 explain 59.7% of the behavior for this model across its entire simulation. Of the 40.3% of behavior not represented in Figure 4, 31.2% of it comes from 469 relatively unimportant feedback loops which individually produce less then 2% of the cumulative model behavior. This leaves 4 remaining feedback loops which are not included. Those 4 loops contribute the remaining 8.9% of cumulative behavior. Those 4 have not been included because they consist of two sets of paired feedback loops (one balancing and one reinforcing) which perfectly cancel each other at all time points, therefore making all 4 of those loops irrelevant to the observed behavior of the model. This overview confirms that Figure 8 is a reasonable simplification of the important dynamics within the Economic Cycles model.

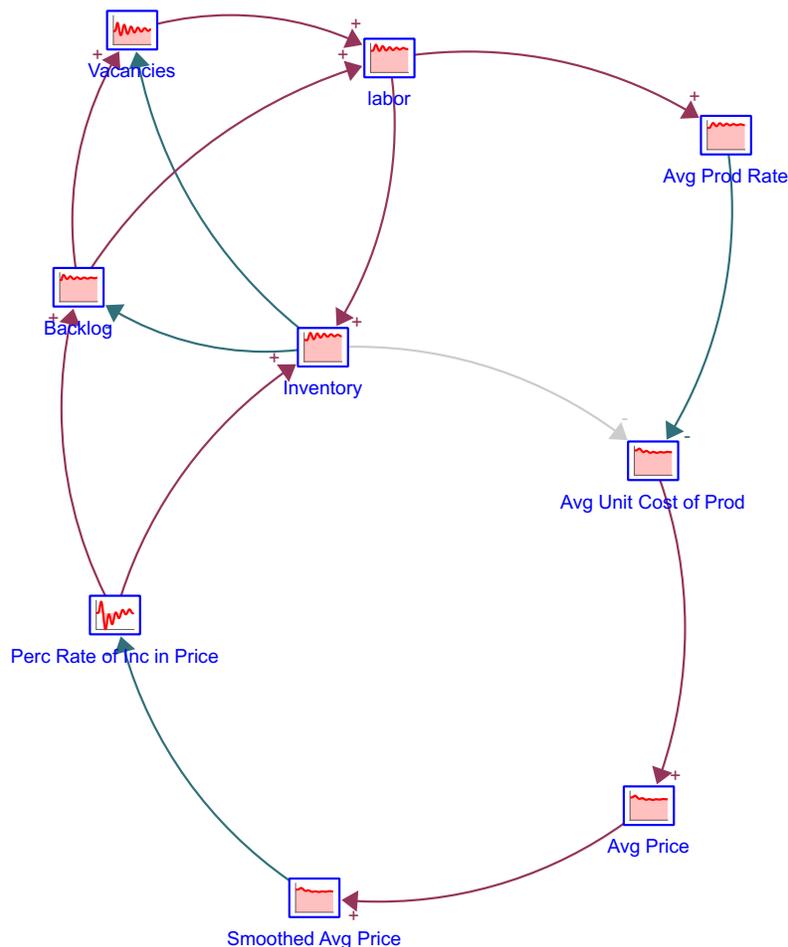

*Figure 8: Automatically generated simplified CLD of Mass' 1975 Economic Cycles model with a link inclusion threshold of over 100%, a loop inclusion threshold of 2.4% and without automatically keeping flows with stocks.*

---

[2] There are 4 other two variable, stock/flow balancing feedback loops, each in their own cycle partition which do not affect the behavior of the model.

*Table 1: List of all feedback loops shown in Figure 4 with cumulative composite link scores. Total loops aggregated describes the number of full feedback loops which are represented by the simplified feedback loop. The total behavior explained by Figure 4 is 59.7% of the total model behavior. B5, B6, R2, and U1 are artifact loops brought forth by the specific combination of feedback loops selected. They were not directly selected for.*

| Loop | Total Contrib. | Total Loops Aggregated | Links Included |
|---|---|---|---|
| B1 | 38.12% | 1 | Vacancies→labor→Inventory→Backlog |
| B2 | 6.60% | 1 | labor→Inventory→Backlog |
| B3 | 5.11% | 3 | Vacancies→labor→Avg Prod Rate→Avg Unit Cost of Prod→Avg Price→Smoothed Avg Price→Perc Rate of Inc in Price→Inventory |
| R1 | 5.11% | 3 | Vacancies→labor→Avg Prod Rate→Avg Unit Cost of Prod→Avg Price→Smoothed Avg Price→Perc Rate of Inc in Price→Backlog |
| B4 | 4.15% | 1 | Vacancies→labor→Inventory→Avg Unit Cost of Prod→Avg Price→Smoothed Avg Price→Perc Rate of Inc in Price→Backlog |
| B5 | 0.37% | 1 | Vacancies→labor→Inventory |
| B6 | 0.13% | 1 | Avg Unit Cost of Prod→Avg Price→Smoothed Avg Price→Perc Rate of Inc in Price→Backlog→labor→Inventory |
| R2 | 0.10 | 3 | Avg Unit Cost of Prod→Avg Price→Smoothed Avg Price→Perc Rate of Inc in Price→Backlog→labor→Avg Prod Rate |
| U1 | 0.02 | 7 | Avg Unit Cost of Prod→Avg Price→Smoothed Avg Price→Perc Rate of Inc in Price→Inventory |

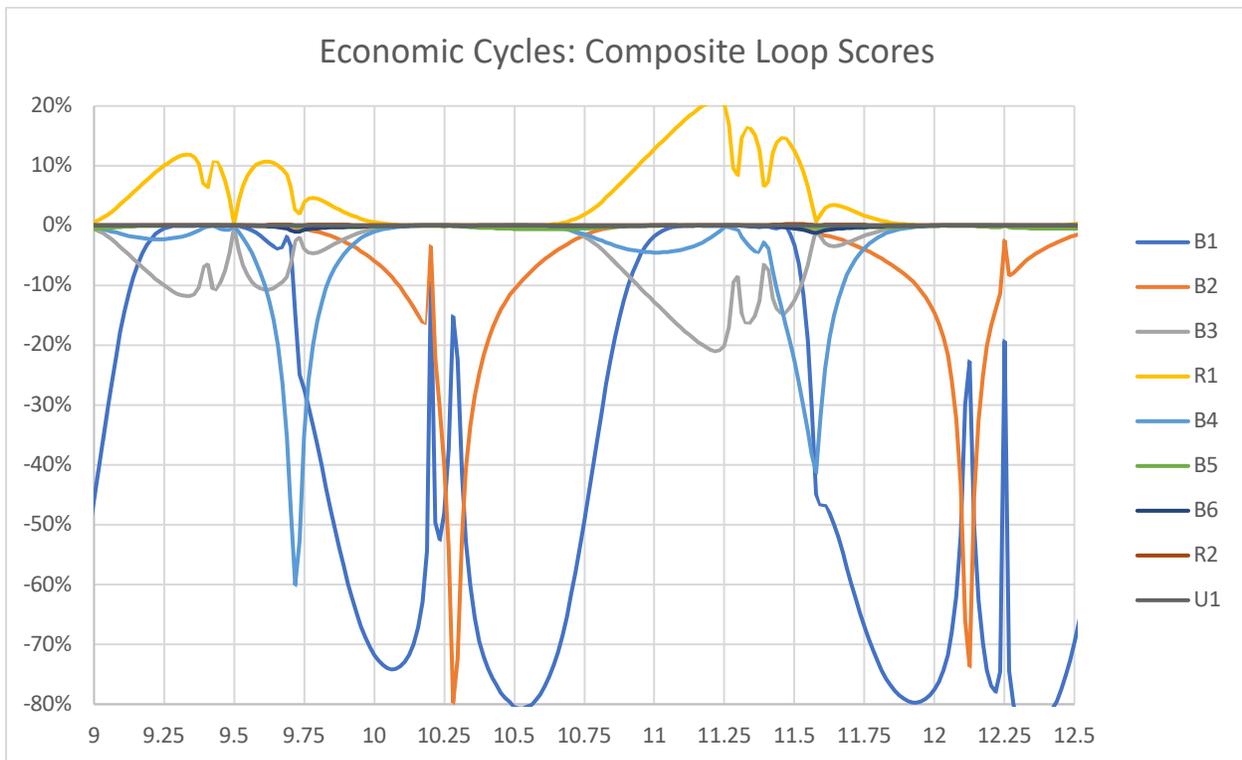

*Figure 9: Composite relative loop scores demonstrating the overtime impact of each of the simplified loops shown in Figure 4 and Table 1. A full explanation of this figure can be seen below.*

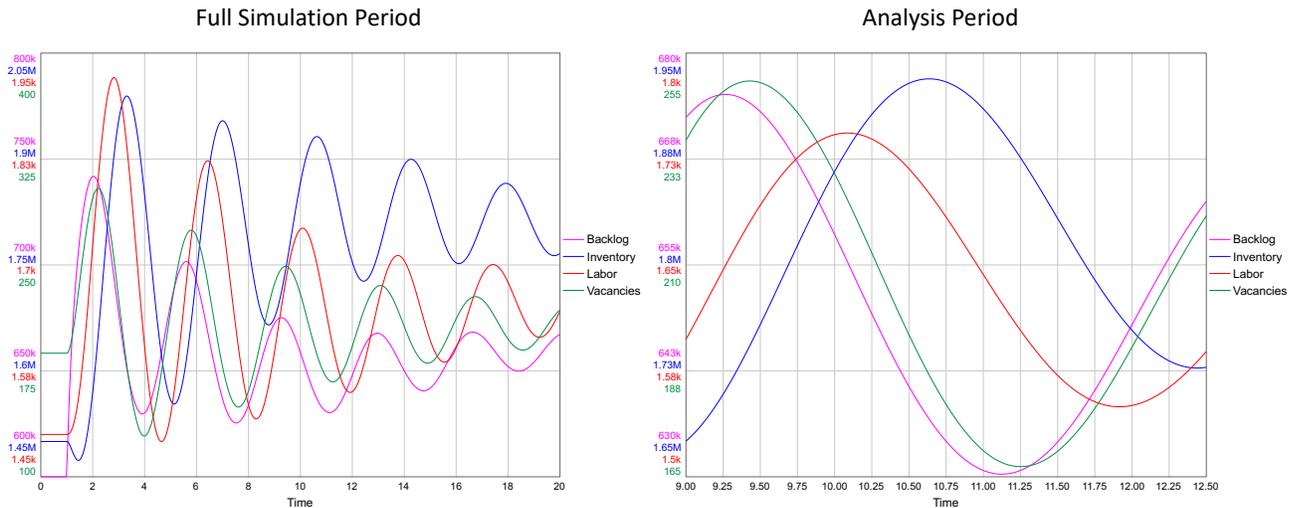

*Figure 10: Plot of the key indicator stocks from Mass's 1975 Economic Cycles model showing the dampened oscillation in economic systems.*

Figure 9 demonstrates from a feedback perspective the dominant relationships at each point in time during a complete single cycle of the model. Figure 10 shows the patterns of behavior for the key stocks in this model over the entire simulation time period, and the chosen period of analysis of Figure 9. The time period of 9 to 12.5 was chosen as the basis for the analysis because it represents one complete cycle of all the stocks in the model. Specifically, this time period tracks one complete wave in Inventory from trough to trough. Within this time period the loop dominance pattern of the Economic Cycle model repeats twice, beginning at time 9 and again at time 10.6. The first walk through this pattern represents the growth of labor, the second, the decline of labor.

The loop dominance pattern within the Economic Cycles model starts with B1 being nearly completely dominant. B1 describes the major drivers of the hiring of labor based upon vacancies created by the backlog driven directly by inventory due to the changes in labor. As the direct hiring process (B1) wanes, a series of loops (R1 and B3) which are both long delays also related to hiring of labor through the changes in vacancies become active. These loops are an example of the perfect destructive interference of feedback loops and cancel each other out perfectly. R1 and B3 cancel each other out because their only difference is the specific route, they take to influencing vacancies from the perceived rate of increase in price. Those two differing pathways taken by R1 and R3 have the same magnitude contribution to vacancies but have opposite signs. R1 represents the effect of the perceived rate of increase in price on backlog which reinforces vacancies, while B3 represents the effect of the same perceived rate of increase in price on inventory which balances vacancies. During this same time period, two other pairs of perfectly destructive feedback loops are also active but are not pictured in Figure 8.

The next major change in the feedback loop dominance of the Economic Cycles model occurs when B4 becomes important. B4, like B1 describes the changes in labor due to vacancies, but it does so through a long delay from the perception of price, due to changes in inventory affecting unit costs, rather than directly from changes in inventory to changes in backlog. After the effect of the delayed price adjustment on vacancies (B4), the direct hiring process becomes dominant again. What is different during this period is that B2 is now active, although not yet dominant. B2 represents the direct effects of the changes in backlog on changes in labor through termination. Next B2 becomes the dominant loop, driving the changes in labor through termination at the inflection points in either the growth or decline of labor. B2 then yields again to B1 which starts the cycle of shifting feedback loop dominance anew. Every other progression through this cycle of feedback loop dominance changes whether labor is growing or shrinking (as B1 is an oscillatory loop), and this progression continues over the entire course of the simulation period. Even as the oscillations dampen, this same progression of feedback loop dominance is still happening.

This cogent explanation for Mass' Economic Cycles model demonstrates the power of the toolset to dramatically simplify the process of model understanding, while providing objective clarity on the causes of behavior in complex systems.

**Demonstrating the application to a discrete system**

Another interesting case is in a much simpler, discrete model, which is ideal for demonstrating the power of Loops that Matter to explain the origins of behavior in discrete systems. Pictured below in Figure 11 is a simple workforce model which uses a conveyor (which can be thought of as a pipeline delay) to model the training process by which apprentices are turned into workers. This system also uses a non-negative stock on purpose to limit the number of employees leaving the system. The equations for this model appear in Table 2.

*Table 2: Equations for the sample workforce system. The two cases analyzed are exactly the same except case 1 used a time to adjust of 5, case 2 used a time to adjust of 2.*

| | |
|---|---|
| *Apprentices = CONVEYOR(hiring – finishing training, training time)* | People |
| *Workers = NONNEGATIVE(finishing training - leaving)* | People |
| *hiring = adjustment + leaving* | People/Time |
| *leaving = 100 + STEP(50, 5)* | People/Time |
| *finishing training = **f** (Apprentices)* | People/Time |
| *adjustment = (target workers – workers)/time to adjust* | People/Time |
| *training time = 5* | Time |
| *target workers = 500* | People |
| *Initial Apprentices = 5\*hiring* | People |
| *Initial Workers = target workers* | People |
| *time to adjust = 5, 2* | Time |

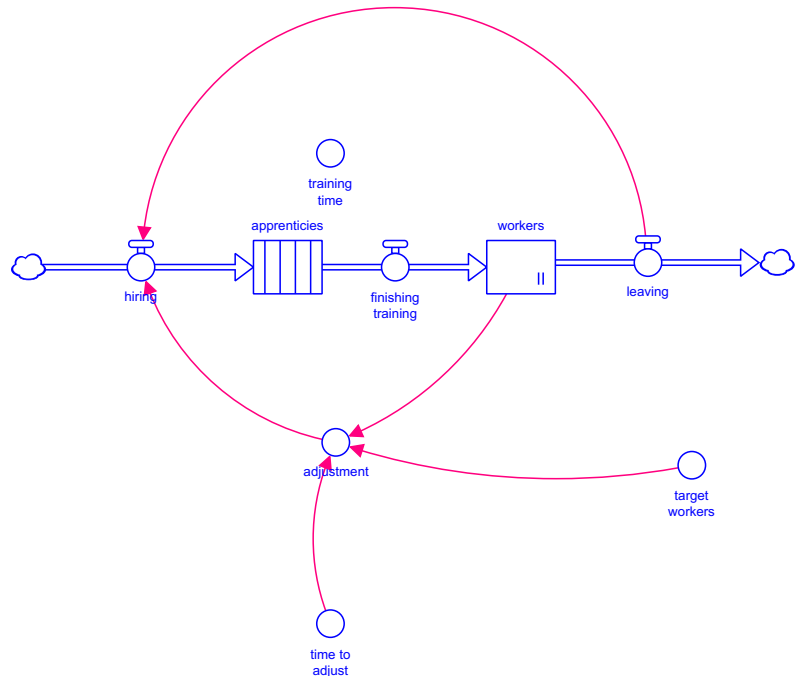

Figure 11: Sample workforce training model containing discrete elements, and non-negative stocks

The simplified CLD seen in Figure 12 highlights the ability of the Loops that Matter method to identify the unique causes of behavior in two different parameterizations of this system highlighting the complexity which is typically hidden in the SFDs of discrete systems. First, Figure 12 shows how Loops that Matter is able to identify the hidden feedback loop (from the perspective of the SFD) between Apprentices and finishing training. It shows how the conveyor is directly affecting its own output, and how changes in the output cause changes to the amount of material in the conveyor. When the time to adjust is equal to 2, the non-negative stock becomes active and constrains the outflow 'leaving', and those dynamics are captured in the CLD on the right in Figure 12. In addition, because 'leaving' now changes, and therefore it changes the number of people hired, it shows up as a concept unto itself in the right CLD. As a matter of fact, the feedback complexity of the model structure changes with its parameterization, the system on the left only has two balancing loops, the system on the right has the same two balancing loops, but it also contains an additional balancing loop (between leaving and workers), and an additional reinforcing loop (across the entirety of the mainchain to adjust hiring without passing through the adjustment variable). These two CLDs in Figure 8 demonstrate that Loops that Matter is fully capable of performing analyses on discrete systems without losing any of its ability to quickly generate insight.

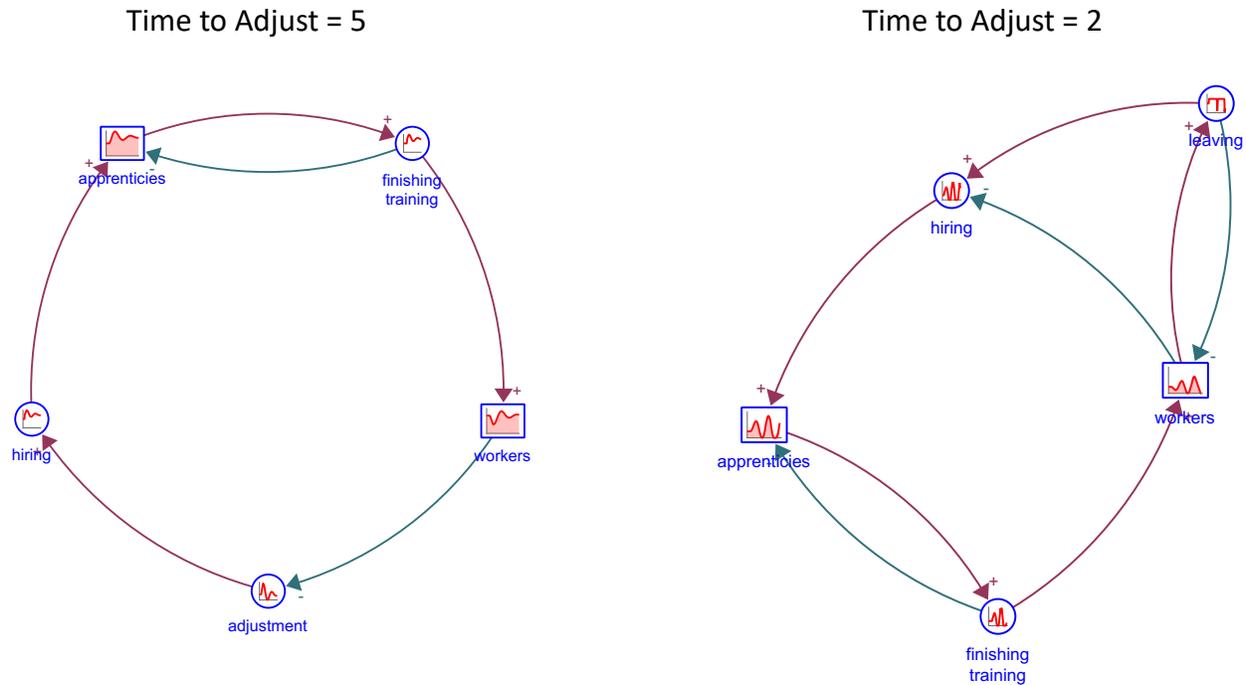

*Figure 12: Causal loop diagrams for the sample workforce training model containing discrete elements under two different parameterizations. Notice how Loops that Matter is able to pick up on the additional feedback loops exposed by the changes in behavior of the discrete system based upon the parameterization of the model. These simplified CLDs communicate more of the feedback dynamics of the system then the SFD pictured in Figure 11.*

**Conclusions**

The new Stella family of products put into the hands of all system dynamics practitioners the automated tools they need to quickly understand the origins of their model's behavior. It makes real the dream, of an objective, correct, useful and easily understandable analysis of the origins of model behavior in any system dynamics models. This paper has demonstrated the utility of the Loops that Matter method, and its visualization capabilities, including animated stock and flow diagrams, animated automatically generated causal loop diagrams, and automated simplification of CLD structure, on large and dynamically complex models such as Mass's Economic Cycles, as well as discrete feedback systems. This research has led us to the point where the only limitations of the Loops that Matter method are its inability to report on unobserved behavioral modes, and its inability to report on loop dominance during equilibrium. Neither of these limitations ought to prevent its acceptance and usage by the field of system dynamics and this should dramatically increase the quality of the understanding of why models produce behavior, while increasing the speed at which these insights are discovered. The work presented here while, designed to fulfill the needs of the system dynamics community, is also directly applicable in a multitude of other fields such as data science, machine learning, and automated intelligence. Future research in these areas will help to broaden the appeal of structural dominance analysis to the wider universe of scientists cutting across fields and disciplines.